\pdfoutput=1 
\documentclass[11pt]{article}
\usepackage{amsmath}
\usepackage{amssymb}
\usepackage{graphicx}
\usepackage{xspace}
\usepackage{workshopsymbols}
\usepackage{siunitx}
\usepackage[margin=1.25in]{geometry}
\usepackage[usenames,dvipsnames]{color}
\usepackage{url}
\usepackage{multirow}
\usepackage{bigstrut}
\usepackage{lineno}
\usepackage{tablefootnote}
\usepackage{todonotes}
\usepackage{subfig}
\usepackage[colorlinks = true,
            linkcolor = blue,
            urlcolor  = blue,
            citecolor = blue,
            anchorcolor = blue]{hyperref}
\usepackage[capitalise]{cleveref}

\usepackage[backend=biber,style=numeric-comp, sorting=none, giveninits=true]{biblatex}

\renewbibmacro{in:}{}

\newcommand\pubnumber{Snowmass Preprint }
\newcommand\pubdate{\today}

\def\Title#1{\begin{center} {\LARGE #1 } \end{center}}
\def\Author#1{\begin{center}{ \sc #1} \end{center}}

\def\Address#1{\begin{center}{ \it #1} \end{center}}

\newcommand\pubblock{\rightline{\begin{tabular}{l} \pubnumber\\
         \pubdate \end{tabular}}}
\newenvironment{Abstract}{\begin{quotation} \begin{center}
                       ABSTRACT
     \end{center}\bigskip  }{\end{quotation}}

\newcount\timehh\newcount\timemm
\timehh=\time
\divide\timehh by 60 \timemm=\time
\newcommand{\dateandtime}{\today \ --
  \ifnum\timehh<10 0\fi\number\timehh:\ifnum\timemm<10 0\fi\number\timemm}

\newcommand\snowmass{\begin{center}\rule[-0.2in]{\hsize}{0.01in}\\\rule{\hsize}{0.01in}\\
\vskip 0.1in Submitted to the  Proceedings of the US Community Study\\ 
on the Future of Particle Physics (Snowmass 2021)\\ 
\rule{\hsize}{0.01in}\\\rule[+0.2in]{\hsize}{0.01in} \end{center}}

\NewDocumentCommand {\AtlasMC} { o m } {%
  \IfNoValueTF {#1} {%
    \textsc{#2}\xspace%
  }{%
    \textsc{#2}\,#1\xspace%
  }%
}
\def\mgamc{\textsc{MadGraph5}\_a\textsc{MC@NLO}\xspace}
\NewDocumentCommand {\PYTHIA} { o } {\AtlasMC[#1]{Pythia}}
\NewDocumentCommand {\WHIZARD} { o } {\AtlasMC[#1]{Whizard}}
\NewDocumentCommand {\CIRCETWO} { o } {\AtlasMC[#1]{Circe2}}
\NewDocumentCommand {\HELAC} { o } {\AtlasMC[#1]{HELAC-Onia}}

\newcommand{\Zd}{\ensuremath{ Z_\text{d}}\xspace} 
\newcommand{\tauleptons}{\ensuremath{\tau}-leptons\xspace}
\newcommand*{\pT}{\ensuremath{p_{\text{T}}}\xspace}
\newcommand{\qbar}{\ensuremath{{\bar q}}}
\newcommand{\fbar}{\ensuremath{{\bar f}}}
\newcommand{\dbar}{\ensuremath{{\bar d}}}
\newcommand{\ubar}{\ensuremath{{\bar u}}}
\newcommand{\sbar}{\ensuremath{{\bar s}}}
\newcommand{\cbar}{\ensuremath{{\bar c}}}
\newcommand{\bbar}{\ensuremath{{\bar b}}}

\begin{document}

\pubblock

\Title{Prospects for searches for Higgs boson decays to dark photons at the ILC}

\bigskip 

\Author{S. Snyder$^1$, C. Weber$^1$, and D. Zhang$^2$}

\medskip

\Address{$^1$Brookhaven National Laboratory, Upton, NY, 11973\\
         $^2$Institute of High Energy Physics, Chinese Academy of Sciences, Shijingshan District, Beijing, 100049}

\medskip

\begin{Abstract}
  An interesting model of dark matter involves a hidden sector decoupled
  from Standard Model (SM) fields except for some portal interaction.
  A concrete realization of this is the Hidden Abelian Higgs Model, which
  gives rise to decays of the SM Higgs boson into a pair of new bosons,
  called $\Zd$ or dark photons.  This note explores prospects for the search
  for such dark photons at the ILC with $\sqrt{s}=\SI{250}{\GeV}$,
  where the dark photons decay promptly.
  For the $H\rightarrow \Zd\Zd\rightarrow 4\ell$
  ($\ell = e,\mu$) final state,
  it follows closely recent similar searches at the LHC, 
  while for the $2\ell2j$ and $4j$ final states a multivariate analysis approach is used.
  This study has not been approved by the SiD consortium.
\end{Abstract}

\snowmass

\def\thefootnote{\fnsymbol{footnote}}
\setcounter{footnote}{0}

\section{Introduction}

A major open question in particle physics is the nature of the
astrophysically-motivated dark matter.  An attractive strategy for
incorporating dark matter into the Standard Model (SM) is through
a hidden sector, decoupled from known SM fields except for some `portal'
interaction~\cite{Fayet:2004bw,Finkbeiner:2007kk,ArkaniHamed:2008qn,Dudas:2012t1,Curtin:2014cca,Curtin:2013fra,Davoudiasl:2013aya,Davoudiasl:2012ag,wells2008find,gopalakrishna2008higgs,Alexander:2016aln}.
A concrete realization of this is the Hidden Abelian
Higgs Model (HAHM)~\cite{Curtin:2014cca,Curtin:2013fra,Davoudiasl:2013aya,Davoudiasl:2012ag,wells2008find,gopalakrishna2008higgs}, in which a new
$U(1)_d$ dark gauge field kinetically mixes with the SM $U(1)_Y$
hypercharge gauge field with some strength $\epsilon$~\cite{Galison:1983pa,Holdom:1985ag,Dienes:1996zr}.
This gives rise to a new Higgs-like dark scalar $S$ along with the
gauge boson of the new field, $\Zd$, or `dark photon'.
The scalar $S$ mixes with the SM Higgs boson with strength $\kappa$,
allowing decays of the SM Higgs boson into pairs of $\Zd$ bosons
via mixing with the $S$ scalar.
For $\epsilon \ll 1$, the decays of the $\Zd$ boson are largely determined
by the gauge couplings.
Over the range $\SI{1}{\GeV} < m_{\Zd} < \SI{60}{GeV}$,
the branching fraction of the $\Zd$~boson into pairs of electrons
or muons could be 10\%--15\%~\cite{Curtin:2014cca}.  These decays
would be prompt for $\epsilon \gtrsim 10^{-5}$.

\begin{figure}
\begin{center}
\includegraphics[width=0.40\hsize]{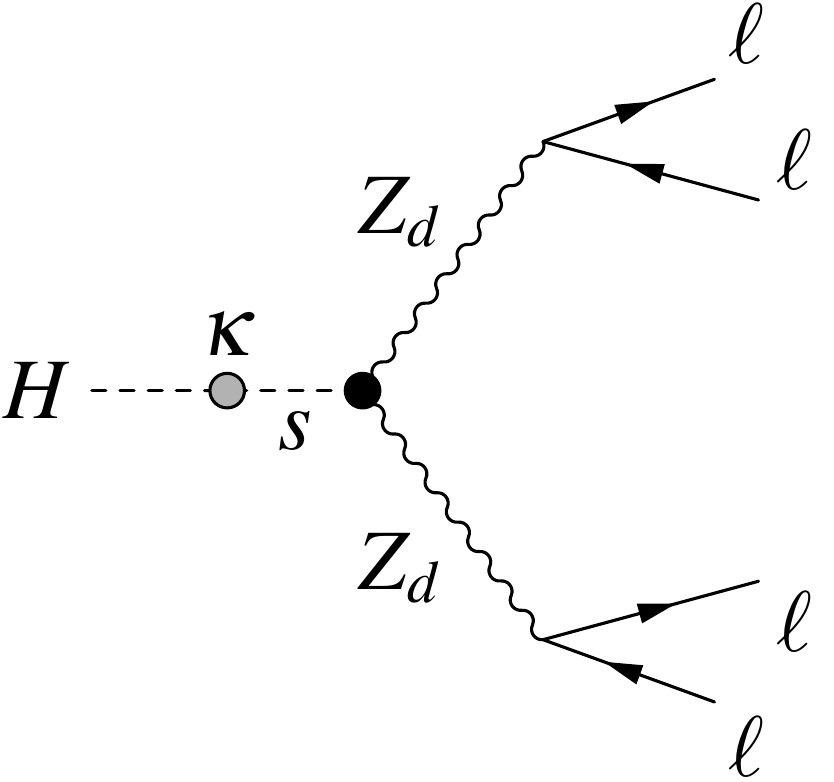}
\end{center}
\caption{Exotic decay of the Higgs~boson into four leptons induced by intermediate
  dark vector bosons via the Higgs portal, where $s$ is a
  dark Higgs boson~\protect\cite{Curtin:2014cca}.
  The $\Zd$ gauge boson decays into SM particles through kinetic mixing
  with the hypercharge field (with branching ratios that are nearly
  independent of $\epsilon$). The
   $H\Zd\Zd$ vertex factor is proportional to
  $\kappa$.}
\label{fig:zd-diag}
\end{figure}

This note explores the prospects for a search for $H\rightarrow \Zd\Zd$
at the ILC with $\sqrt{s}=\SI{250}{\GeV}$, with the $\Zd$ bosons decaying
promptly to $4\ell$ ($\ell\equiv e,\mu$),
as illustrated in \cref{fig:zd-diag}, as well as $2\ell2j$ and $4j$.  
For the $4\ell$ final state, it follows closely the analysis
of ATLAS at the LHC with $\sqrt{s}=\SI{13}{\TeV}$ and an integrated
luminosity of $\SI{139}{\ifb}$~\cite{HDBS-2018-55}.
Other similar searches, including searches for pairs of light bosons decaying
into muons, \tauleptons, photons, and/or jets,
as well as searches for a single light boson decaying into a pair of muons,
using both
$\sqrt{s}=\SI{8}{\TeV}$ and \SI{13}{\TeV} data, have been performed by
ATLAS~\cite{HDBS-2018-47,HIGG-2017-09,HIGG-2017-05,HIGG-2016-03,HIGG-2014-02},
CMS~\cite{CMS-HIG-13-010, CMS-HIG-16-015,CMS-HIG-18-024,CMS-EXO-20-014},
and LHCb~\cite{LHCB-PAPER-2017-038}.
Searches for long-lived signatures at ATLAS and CMS are reported
in Refs.~\cite{EXOT-2013-22,EXOT-2014-09,SUSY-2014-02,CMS-EXO-12-037,EXOT-2017-03,SUSY-2017-04,EXOT-2017-28,EXOT-2017-32,CMS-HIG-18-003},
while further searches for a SM Higgs~boson decaying into undetected particles
are reported in Refs.~\cite{HIGG-2018-54,CMS-HIG-17-023}.

The following section describes how the event samples used for these analyses
were simulated and reconstructed.  This is followed by a description of the
$H\rightarrow \Zd\Zd\rightarrow 4\ell$, $2\ell2j$, and $4j$
analyses, including the event selection,
a discussion of quarkonia backgrounds, and expected limits.
The note ends with a summary and discussion of possible future work.

\section{Event simulation and reconstruction}
\label{sec:simulation}

The $ee\rightarrow ZH\rightarrow ff\Zd\Zd$ signal was generated
according to the HAHM~\cite{wells2008find,gopalakrishna2008higgs,Curtin:2013fra, Curtin:2014cca}
implementation in \mgamc\ version 2.8.2~\cite{Alwall:2014hca},
with the Higgs~boson mass set to $m_H = \SI{125}{\GeV}$
and $\epsilon$ and $\kappa$ both set to $10^{-4}$.
The $e^-$ beam polarization was set to $-80\%$ and $e^+$ to $+30\%$.
Leptonically decaying $\Zd$~bosons were forced to decay to either a $ee$ or $\mu\mu$ pair.
Final states with $\tau$-leptons were not included.  In the similar
LHC analysis~\cite{HDBS-2018-55}, the change in signal region yield
due to the omission of these decays was below~1\% and thus neglected. 
Decays of $\Zd$~bosons to jets were allowed to be inclusive.
Other Higgs~boson production mechanisms, which are much smaller
at the ILC with $\sqrt{s}=\SI{250}{\GeV}$, were also omitted.
Showering was performed via \mgamc's built-in interface
to \PYTHIA\ 8.244~\cite{Sjostrand:2014zea}.
Events in the $4\ell$ final state were generated at $m_{\Zd}=\SI{1}{\GeV}$,
every $\SI{2}{\GeV}$ in the range
$\SI{2}{\GeV} \le m_{\Zd} \le \SI{12}{\GeV}$,
and every $\SI{5}{\GeV}$ in the range
$\SI{15}{\GeV} \le m_{\Zd} \le \SI{60}{\GeV}$. 
Signal samples with $2\ell2j$ and $4j$ final states were produced at 
$mZd = \SI{20}{\GeV}$, \SI{40}{\GeV}, and \SI{60}{\GeV}.
Each signal sample was generated with 20,000 events
 per $m_{\Zd}$ value and per final state.

The $ee\rightarrow X+H \rightarrow X+4\ell$ background was generated
using \WHIZARD\ 2.8.5~\cite{Kilian:2007gr,Moretti:2001zz} along with its internal version of
\PYTHIA\ 6.4~\cite{Sjostrand:2006za}, with parameters set corresponding
to the OPAL tune provided in~\cite{ilc-walkthrough}.
The $e^-$ beam polarization was set to $-80\%$ and $e^+$ to $+30\%$,
and the provided \verb|ilc250| \CIRCETWO\ parameterization was used.
The Higgs~boson was forced to decay into either a $4e$, $2e2\mu$,
or $4\mu$ final state.  For this background, 20,000 events were generated.

The non-resonant $ee\rightarrow 4e, 2e2\mu, 4\mu$ background was also
generated with \WHIZARD, using the same settings as described previously.
To remove divergences, the final state leptons were restricted
to be within $\SI{20}{\degree} < \theta < \SI{160}{\degree}$;
further, each pair of final
state leptons had to have an invariant mass greater than $\SI{1}{\GeV}$.
For this background, 250,000 events were generated.

The generated events were processed through the full iLCSoft simulation
and reconstruction chain~\cite{ilcsoft} using the \verb|o2_V03| version
of the SiD~\cite{ilc-tdr-v4} geometry.  The electrons and muons
used in the analysis were identified by
\verb|IsolatedLeptonTaggingProcessor|~\cite{iltp}.
This, however, relies on a cone-based isolation algorithm that rejects
leptons from $\Zd\rightarrow\ell\ell$ decay if
$m_{\Zd} \lesssim \SI{4}{\GeV}$.  In order extend the analysis
to smaller values of $m_{\Zd}$, the isolation algorithm is modified.
When summing the particle flow objects inside an isolation cone around
an electron or muon candidate, the highest-energy same-flavor
opposite-sign lepton candidate is ignored.  This recovers
efficiency for $\Zd\rightarrow\ell\ell$ at low $m_{\Zd}$
with no significant increase in background for this analysis
(as estimated from a $b\bbar$ sample generated with \WHIZARD).

The $2\ell2j$ and $4j$ final states rely on inclusive $ee\rightarrow ZH$ 
and $ee\rightarrow ZZ$ samples from the SiD collaboration~\cite{potter2021sid}. 
These were generated using \WHIZARD\ 2.6.4 using nominal ILC Technical Design Report 
polarization fractions of 80\% polarized electrons and 30\% polarized positrons. 
However the study is limited to samples with $-80\%$ and $e^+$ to $+30\%$ 
polarization to match electron beam polarization of the signal samples, 
and has not been approved by the SiD consortium.

\section{\texorpdfstring{$H\rightarrow\Zd\Zd\rightarrow 4\ell$}{H->ZdZd->4l} event selection}

The event selection closely follows that of the
ATLAS $H\rightarrow\Zd\Zd\rightarrow 4\ell$ analysis~\cite{HDBS-2018-55},
with a few modifications for the ILC environment, such as removing
detector-specific particle-identification requirements and
changing requirements on $\pT$ and $\eta$ to $E$ and $\theta$.
Although each signal event contains a $Z$~boson decay in addition
to the Higgs~boson decay, no explicit requirements are made
for the $Z$~boson decay.
Two sets of selections are used.  The first, called the high-mass (HM)
selection is designed for $\SI{15}{\GeV} < m_{\Zd} < \SI{60}{\GeV}$,
and the second, low-mass (LM), selection is designed for
$\SI{1}{\GeV} < m_{\Zd} < \SI{15}{\GeV}$.  In this latter region,
the angular separation between the two leptons from
$\Zd\rightarrow \ell\ell$ decay becomes small.
For the ATLAS analysis~\cite{HDBS-2018-55}, the LM analysis used
only the $4\mu$ final state, due to a decreased efficiency for identifying
closely-spaced electrons.  Simulated ILC events do not show such a drop
in efficiency, so here all final states are used
for the LM analysis.

Electrons and muons must satisfy
$E > \SI{7}{\GeV}$ and $0.35 < \theta < \pi - 0.35$.  Such leptons are
formed into quadruplets consisting of two same-flavor opposite-sign (SFOS)
lepton pairs, giving $4e$, $2e2\mu$, and $4\mu$ final states.
If there are more than two such pairs, multiple quadruplets are formed
from all possible SFOS combinations.  The invariant masses
of the two pairs are denoted by $m_{12}$ and $m_{34}$, with $m_{12}$
taken to be the one closest to the mass of the
$Z$~boson: $|m_{12} < m_Z| < |m_{34} - m_Z|$.

If all four leptons in a quadruplet have the same flavor, then one
can also define alternate pairings.  The invariant mass $m_{14}$
is defined from the positively charged lepton of the $m_{12}$ pair
and the negatively charged lepton of the $m_{34}$ pair.
The other alternative pairing $m_{23}$ is defined similarly.

For all quadruplets, the three highest-energy leptons must satisfy
$E_1 > \SI{20}{\GeV}$, $E_2 > \SI{15}{\GeV}$, and $E_3 > \SI{10}{\GeV}$.
For the HM analysis only, 
the angular separation between all same-flavor leptons must satisfy
$\Delta R(\ell,\ell') > 0.1$ and for different-flavor leptons
$\Delta R(\ell,\ell') > 0.2$,
where $(\Delta R)^2 = (\Delta\eta)^2 + (\Delta\phi)^2$
and $\eta$ is the pseudorapidity\footnote{Although an analysis
  in an $e^-e^+$ environment would more naturally use $\theta$ than $\eta$,
  the use of $\Delta R$ is retained here for consistency with
  the ATLAS analysis.}.
Events are required to have at least one such quadruplet.
If there is more than one,
the quadruplet with the smallest $\Delta m_{\ell\ell} = |m_{12} - m_{34}|$
is used.

For the HM event selection, the invariant mass of the four leptons must
be consistent with that of the SM~Higgs boson:
$\SI{115}{\GeV} < m_{4\ell} < \SI{130}{\GeV}$.  The quadruplet must
also not be consistent with the decay of $Z$~bosons ($Z$-veto):
$\SI{10}{\GeV} < m_{12,34} < \SI{64}{\GeV}$.  For the $4e$
and $4\mu$ channels, it is possible that the leptons are mispaired,
so for these channels there is also a requirement on the
alternative lepton pairings:
$\SI{5}{\GeV} < m_{14,23} < \SI{75}{\GeV}$.
Events with lepton pairs consistent with $J/\psi$ or $\Upsilon$ decay
are also rejected.  Events are rejected if any of $m_{12,34,14,23}$
are in the ranges
$(m_{J/\psi} - \SI{0.25}{\GeV})$ to $(m_{\psi(2S)} + \SI{0.30}{\GeV})$ or
$(m_{\Upsilon(2S)} - \SI{0.70}{\GeV})$ to $(m_{\Upsilon(3S)} + \SI{0.75}{\GeV})$,
where the quarkonia masses are taken to be
$m_{J/\psi} = \SI{3.096}{\GeV}$, $m_{\psi(2S)} = \SI{3.686}{\GeV}$,
$m_{\Upsilon(1S)} = \SI{9.461}{\GeV}$,
and $m_{\Upsilon(3S)} = \SI{10.355}{\GeV}$~\cite{pdg20}.
Finally, a requirement $m_{34} / m_{12} > 0.85$ ensures that the
two pairs have similar invariant masses.

For the LM event selection, the pair invariant masses are required
to be in the range $\SI{0.8}{\GeV} < m_{12,34} < \SI{20}{\GeV}$.
The requirement on the overall invariant mass is tightened to
$\SI{120}{\GeV} < m_{4\ell} < \SI{130}{\GeV}$ due to smaller
radiative tails in this regime.  The $Z$-veto requirement is
not applied, and and only the alternate lepton pairings are used
for the quarkonia vetoes (see \cref{sec:4l-onia}).
The final requirement $m_{34} / m_{12} > 0.85$ is the same as for the
HM analysis.

Both event selections are summarized in \cref{tab:4lselection}.

\begin{table}
\caption{Summary of event selection requirements for the HM and LM
  analyses.  
}
\centering
\small
\vskip 1em
\begin{tabular}{|c|c|>{\centering}m{5cm}|>{\centering}m{5cm}|}
\cline{3-4} \cline{4-4}
\multicolumn{1}{c}{} &
  & High-mass (HM) analysis \bigstrut[t]\\
    $H\rightarrow \Zd\Zd\rightarrow4\ell$ $(\ell=e,\mu)$ \bigstrut[b]
  & Low-mass (LM) analysis\\
    $H\rightarrow \Zd\Zd\rightarrow4\ell$ $(\ell=e,\mu)$ \tabularnewline
\hline

\multicolumn{2}{|c|}{Mass range\bigstrut}
  & $\SI{15}{\GeV} < m_X < \SI{60}{\GeV}$ \bigstrut
  & $\SI{1}{\GeV} < m_X < \SI{15}{\GeV}$ \bigstrut\tabularnewline
\hline

\multicolumn{2}{|c|}{Leptons}
  & \multicolumn{2}{>{\centering}m{10cm}|}{Four isolated electrons or muons\bigstrut[t]\\
  with $E > \SI{7}{\GeV}$ and $0.35 < \theta < \pi - 0.35$\bigstrut[b]}
  \tabularnewline
\hline

\multicolumn{2}{|c|}{Quadruplet selection}
  & \multicolumn{2}{>{\centering}m{10cm}|}{$e^+e^-e^+e^-$, $e^+e^-\mu^+\mu^-$, or $\mu^+\mu^-\mu^+\mu^-$;\bigstrut[t] \\
    Three leading-$E$ leptons satisfying $E > \SI{20}{\GeV}$, $\SI{15}{\GeV}$, $\SI{10}{\GeV}$\\
    Define pairs $m_{12}$ and $m_{34}$ such that $|m_{12}-m_Z| < |m_{34} - m_Z|$  \bigstrut[b]} \tabularnewline
\cline{3-4}

\multicolumn{1}{|c}{}  &
  & \multicolumn{1}{>{\centering}m{5cm}|}{\bigstrut[t]$\Delta R(\ell,\ell^{\prime})>0.10\,(0.20)$ for same-flavor (different-flavor) $\ell,\ell^{\prime}$\bigstrut}
  & ---\bigstrut \tabularnewline
\hline

\multicolumn{2}{|c|}{Quadruplet ranking} \bigstrut
  & \multicolumn{2}{>{\centering}m{10cm}|}{Select quadruplet with smallest $\Delta m_{\ell\ell}=|m_{12}-m_{34}|$}\tabularnewline
\hline

\multirow{10}{*}{\parbox{1.5cm}{Event\\ selection}} 
 & $m_{4\ell}$ \bigstrut
  & $\SI{115}{\GeV} < m_{4\ell} < \SI{130}{\GeV}$
  & $\SI{120}{\GeV} < m_{4\ell} < \SI{130}{\GeV}$ \tabularnewline
\cline{2-4}

 & $Z$-veto
  & $\SI{10}{\GeV} < m_{12,34} < \SI{64}{\GeV}$\bigstrut[t] \\
    For $4e$ and $4\mu$ channels: $\SI{5}{\GeV} < m_{14,23} < \SI{75}{\GeV}$ \bigstrut[b]
   & ---
  \tabularnewline
     \cline{2-4}

 & Heavy-flavor veto
  & \multicolumn{1}{>{\centering}m{5cm}|}{Reject event if $m_{12,34,14,23}$ in:\bigstrut[t]\\
  $(m_{J/\psi}-\SI{0.25}{\GeV})$ to $(m_{\psi(2S)}+\SI{0.30}{\GeV})$, or\\
  $(m_{\Upsilon(1S)}-\SI{0.70}{\GeV})$ to $(m_{\Upsilon(3S)}+\SI{0.75}{\GeV})$}\bigstrut[b]
  & \multicolumn{1}{>{\centering}m{5cm}|}{Reject event if $m_{14,23}$ in:\bigstrut[t]\\
  $(m_{J/\psi}-\SI{0.25}{\GeV})$ to $(m_{\psi(2S)}+\SI{0.30}{\GeV})$, or\\
  $(m_{\Upsilon(1S)}-\SI{0.70}{\GeV})$ to $(m_{\Upsilon(3S)}+\SI{0.75}{\GeV})$}\bigstrut[b]
\tabularnewline
\cline{2-4}

 & Signal region\bigstrut
  & $m_{34}/m_{12} > 0.85$
  & $\SI{0.8}{\GeV} < m_{12,34} < \SI{20}{\GeV}$\bigstrut[t]\\
  $m_{34}/m_{12}>0.85$
  \bigstrut[b]
    \tabularnewline
  \hline
\end{tabular}
\label{tab:4lselection}
\end{table}

The main backgrounds are from SM $H\rightarrow ZZ^*\rightarrow 4\ell$ decay
and also from nonresonant $4\ell$ production, with the latter source dominating.
These are estimated
using the simulated samples described in~\cref{sec:simulation}.
Backgrounds in which jets are misidentified as leptons are assumed
to be negligible.  Estimated backgrounds for a data sample
of $\SI{2000}{\ifb}$ are shown in \cref{tab:4lbackgrounds}.

\begin{table}
  \caption{Estimated backgrounds for the $H\rightarrow \Zd\Zd\rightarrow 4\ell$
    analyses for a data sample of $\SI{2000}{\ifb}$.
    No simulated events for the non-resonant background pass the LM selection,
    so a limit is given based on the background represented by
    one simulated event.}
\centering
\small
\vskip 1em
\begin{tabular}{lcccc}
\hline
HM selection\bigstrut                & $4e$           & $2e2\mu$& $4\mu$         & All \\
\hline
\bigstrut[t]
$H\rightarrow ZZ^*\rightarrow 4\ell$ & $0.37 \pm 0.04$& $0.39 \pm 0.04$& $0.46 \pm 0.04$& $1.22 \pm 0.07$\\
Non-resonant                         & $3.05 \pm 1.15$& $2.62 \pm 1.07$& $0.44 \pm 0.44$& $6.10 \pm 1.63$\\
\bigstrut[b]
Total                                & $3.42 \pm 1.15$& $3.00 \pm 1.07$& $0.90 \pm 0.44$& $7.32 \pm 1.63$\\
\hline
\\
\hline
LM selection\bigstrut                & $4e$           & $2e2\mu$& $4\mu$         & All \\
\hline
\bigstrut[t]
$H\rightarrow ZZ^*\rightarrow 4\ell$ & $0.04 \pm 0.01$& $< 0.01$& $0.03 \pm 0.01$& $0.07 \pm 0.02$\\
Non-resonant                         &   $< 0.44$&  $< 0.44$&         $< 0.44$&         $< 0.44$\\
\bigstrut[b]
Total                                &  $< 0.47$&  $< 0.44$&         $< 0.47$&         $< 0.51$\\
\hline

\end{tabular}
\label{tab:4lbackgrounds}
\end{table}

\section{Quarkonia backgrounds to \texorpdfstring{$H\rightarrow\Zd\Zd\rightarrow 4\ell$}{H->ZdZd->4l} final states}
\label{sec:4l-onia}

The ATLAS $H\rightarrow \Zd\Zd\rightarrow 4\ell$ analysis~\cite{HDBS-2018-55}
has no sensitivity in the regions $\SI{2}{\GeV} < m_{\Zd} < \SI{4.4}{\GeV}$
and $\SI{8}{\GeV} < m_{\Zd} < \SI{12}{\GeV}$ due to the presence of large
backgrounds from quarkonia production.  However, at the ILC, hadronic
backgrounds such as this will be much smaller, and the excellent lepton
momentum resolution of the ILC detectors may reduce the size of the
$m_{\Zd}$ ranges affected by these backgrounds.  Unfortunately, there is
no reliable, general-purpose simulation of quarkonia production at the ILC.
But one can still estimate these backgrounds, as described below.

First, consider direct, non-resonant production of $J/\psi$ and $\Upsilon$
pairs.  These processes were estimated using
\HELAC version 2.0.1~\cite{Shao:2015vga, Shao:2012iz}.
This can calculate processes such as
$e^-e^+\rightarrow J/\psi J/\psi$ and
$e^-e^+\rightarrow J/\psi J/\psi + q\qbar$,
and similarly for $\Upsilon$.  The quarkonia can be produced as either
color singlets or color octets.  For the purpose of this study,
excited quarkonia states are not considered.  The initial beams
are set to $e^-e^+$ with beam energies of $\SI{125}{\GeV}$ each,
with initial-state radiation disabled.  (\HELAC does not implement
beam polarization effects, and enabling initial-state radiation
caused the subsequent showering step to fail.)  For the processes
with the largest cross sections, generated events
were then showered and hadronized with \PYTHIA\ 8.244,
with the quarkonia forced to decay to either $e^-e^+$
or $\mu^-\mu^+$.  The quarkonia decay branching ratios were taken
to be $\BR(J/\psi\rightarrow \ell\ell) = 0.119$
and $\BR(\Upsilon\rightarrow \ell\ell) = 0.049$~\cite{pdg20}.
Events were then passed through the detector simulation and analysis,
yielding estimates of these quarkonia backgrounds after the
LM selection, shown in~\cref{tab:4l-direct-onia}.  These estimates are
all much smaller than other backgrounds for this selection.

\begin{table}
\def\cs{^{{\bf 1}}}
\def\co{^{{\bf 8}}}
  \caption{Quarkonia production processes calculated with \HELAC.
    The $\cs$ and $\co$ superscripts denote color-singlet and
    color-octet states, respectively.  For the processes with larger
    cross sections, the number of expected events passing the LM selection
    for a data sample of \SI{2000}{\ifb} is shown.  This is shown
    as a limit for cases where no simulated events pass the selection.}
\vskip 1em
\begin{minipage}{\linewidth}
\small
\centering
\begin{tabular}{lSS}
  \hline
  \bigstrut
  Process & {Cross section (ab)} & {Expected background} \\
  \hline
  \bigstrut[t]
  $J/\psi\cs J/\psi\cs$          & 8.3    & < 5.7e-6 \\
  $J/\psi\co J/\psi\co$          & 2.9e-6 \\
  $J/\psi\cs J/\psi\cs + d\dbar$ & 0.0033 \\
  $J/\psi\co J/\psi\co + d\dbar$ & 11     & < 0.0018 \\
  $J/\psi\cs J/\psi\cs + u\ubar$ & 0.0017 \\
  $J/\psi\co J/\psi\co + u\ubar$ & 45     & < 0.0034 \\
  $J/\psi\cs J/\psi\cs + s\sbar$ & 0.0033 \\
  $J/\psi\co J/\psi\co + s\sbar$ & 11     & < 0.0018 \\
  $J/\psi\cs J/\psi\cs + c\cbar$ & 0.057 \\
  $J/\psi\co J/\psi\co + c\cbar$ & 41     & 6.3e-4 \\
  $J/\psi\cs J/\psi\cs + b\bbar$ & 0.0022 \\
  $J/\psi\co J/\psi\co + b\bbar$ & 8.3    & 1.0e-4 \\
  $J/\psi\cs J/\psi\cs + gg$     & 0.012 \\
  $J/\psi\co J/\psi\co + gg$     & 1.6    & {\footnote{\HELAC failed to generate events for this process.}}\\
  \bigstrut[b]
  $J/\psi\co J/\psi\co + g$      & 0.060  & 1.1e-6 \\
  \hline
  \bigstrut[t]
  $\Upsilon\cs \Upsilon\cs$          & 0.18  & < 1.3e-7 \\
  $\Upsilon\co \Upsilon\co$          & 6.5e-11 \\
  $\Upsilon\cs \Upsilon\cs + d\dbar$ & 8.1e-4 \\
  $\Upsilon\co \Upsilon\co + d\dbar$ & 1.0e-5 \\
  $\Upsilon\cs \Upsilon\cs + u\ubar$ & 0.0034 \\
  $\Upsilon\co \Upsilon\co + u\ubar$ & 4.0e-5 \\
  $\Upsilon\cs \Upsilon\cs + s\sbar$ & 8.1e-4 \\
  $\Upsilon\co \Upsilon\co + s\sbar$ & 1.0e-5 \\
  $\Upsilon\cs \Upsilon\cs + c\cbar$ & 0.0031 \\
  $\Upsilon\co \Upsilon\co + c\cbar$ & 3.9e-5 \\
  $\Upsilon\cs \Upsilon\cs + b\bbar$ & 0.0022 \\
  $\Upsilon\co \Upsilon\co + b\bbar$ & 8.4e-6 \\
  $\Upsilon\cs \Upsilon\cs + gg$     & 4.7e-5 \\
  $\Upsilon\co \Upsilon\co + gg$     & 8.0e-5 \\
  \bigstrut[b]
  $\Upsilon\cs \Upsilon\cs + g$      & 2.9e-7 \\
  \hline
\end{tabular}
\end{minipage}
\label{tab:4l-direct-onia}
\end{table}

Another possibility is the decay of a $Z$~boson into a quarkonia pair,
or two $Z$~bosons decaying to the same quarkonium state.
Backgrounds from these processes were estimated using current
experimental limits/measurements for such decays~\cite{pdg20}:
\begin{center}
  \begin{tabular}{lS}
    $\BR(Z\rightarrow J/\psi + X)$ & 5.1e-3 \\
    $\BR(Z\rightarrow J/\psi J/\psi)$ & < 2.2e-6 \\
    $\BR(Z\rightarrow \Upsilon + X)$ & 1.0e-4 \\
    $\BR(Z\rightarrow \Upsilon\Upsilon)$ & < 1.5e-6 \\
  \end{tabular}
\end{center}
\WHIZARD was used to generate a sample of $e^-e^+\rightarrow f\fbar Z$,
where $f$ is any fermion, configured as described in \cref{sec:simulation},
except that the \PYTHIA showering was set to force the $Z$~boson to decay
as either $Z\rightarrow J/\psi J/\psi\rightarrow 4\ell$ or
$Z\rightarrow \Upsilon\Upsilon \rightarrow 4\ell$.  Events were then
passed through the detector simulation and LM selection.
The total cross section calculated by \WHIZARD for $f\fbar Z$
was \SI{7.1e3}{fb}.  Taking into account the branching ratios above
and the efficiency of the LM selection, the expected background
for a \SI{2000}{\ifb} data sample is $< \SI{2.2e-5}{}$ for
$Z\rightarrow J/\psi J/\psi$ and \SI{1.2e-5}{} for
$Z\rightarrow \Upsilon\Upsilon$.

Similarly, \WHIZARD was also used to generate a sample of
$e^-e^+\rightarrow ZZ$, where here each $Z$~boson was forced to decay
as either $Z\rightarrow J/\psi c\cbar$ or $Z\rightarrow \Upsilon b\bbar$.
The total \WHIZARD cross section for this process was \SI{1.8e3}{fb}.
Again, taking into account branching ratios and efficiencies,
the expected background
for a \SI{2000}{\ifb} data sample is 0.0029 for
$ZZ\rightarrow J/\psi J/\psi + c\cbar c\cbar$ and \SI{4.3e-7}{} for
$ZZ\rightarrow \Upsilon\Upsilon + b\bbar b\bbar$.

A final possibility is $H\rightarrow J/\psi J/\psi$ or
$H\rightarrow \Upsilon\Upsilon$.  Such a background would be particularly
concerning since it could not be removed by the requirement that the
overall invariant mass be consistent with that of the SM Higgs~boson.
However, calculations give
$BR(H\rightarrow J/\psi J/\psi) = \SI{5.9e-10}{}$ and
$BR(H\rightarrow \Upsilon\Upsilon) = \SI{4.3e-10}{}$~\cite{Gao:2022iam,Kartvelishvili:2008tz},
which are much smaller than the $H\rightarrow\Zd\Zd$ branching ratio to which
the LM analysis is sensitive.

Although this does not exhaust all possibilities for quarkonia
background processes, it should be a representative sample.  All processes
examined result in backgrounds that are much smaller than the other
(already-small) backgrounds to the LM selection.  Therefore, quarkonia
production is unlikely to be a significant background to this analysis.
However, evaluating this with more confidence would likely require improved
codes for calculating quarkonia processes.

\begin{figure}
\begin{center}
  \subfloat[]{
    \includegraphics[width=0.49\textwidth]{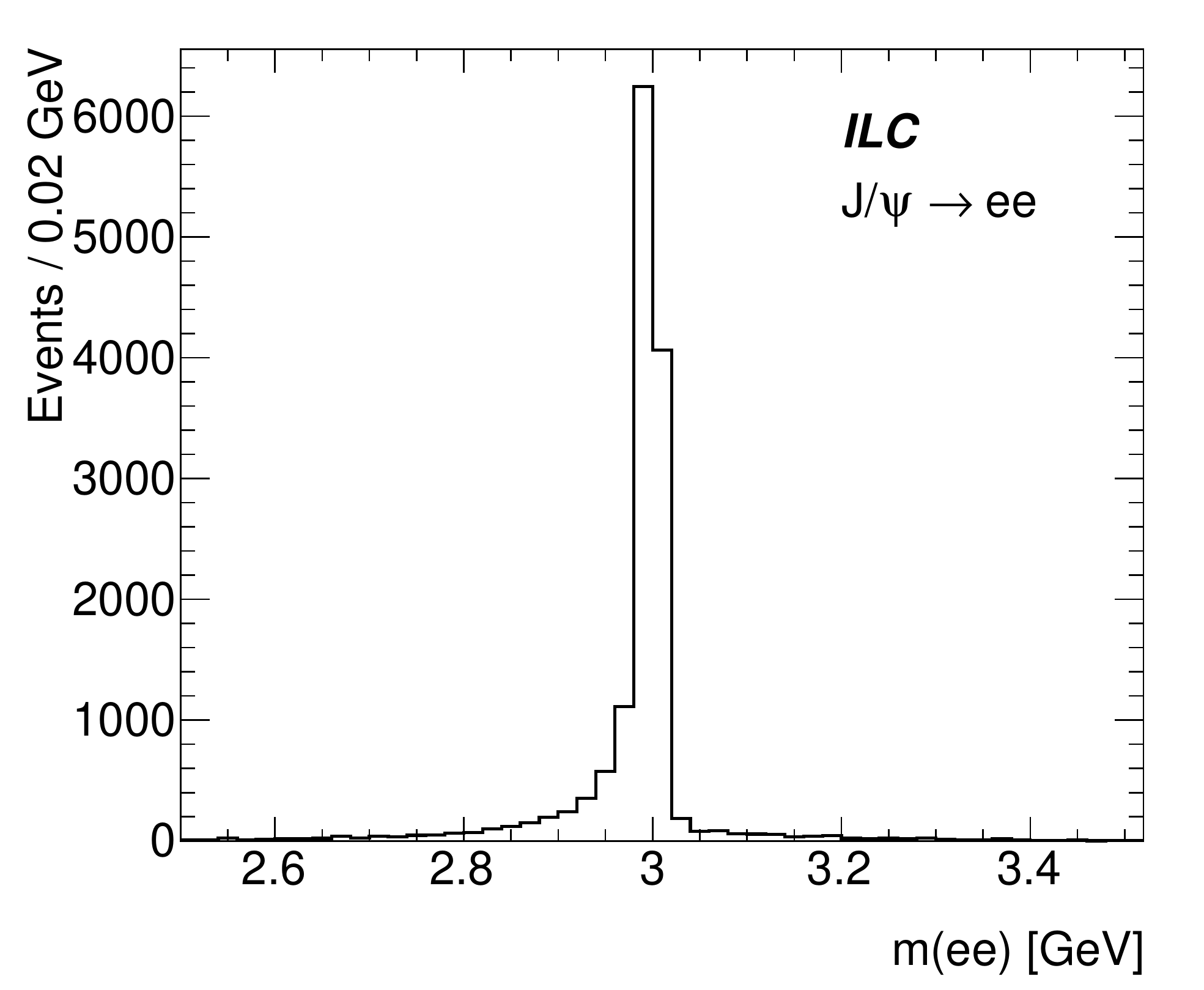}}
  \subfloat[]{
    \includegraphics[width=0.49\textwidth]{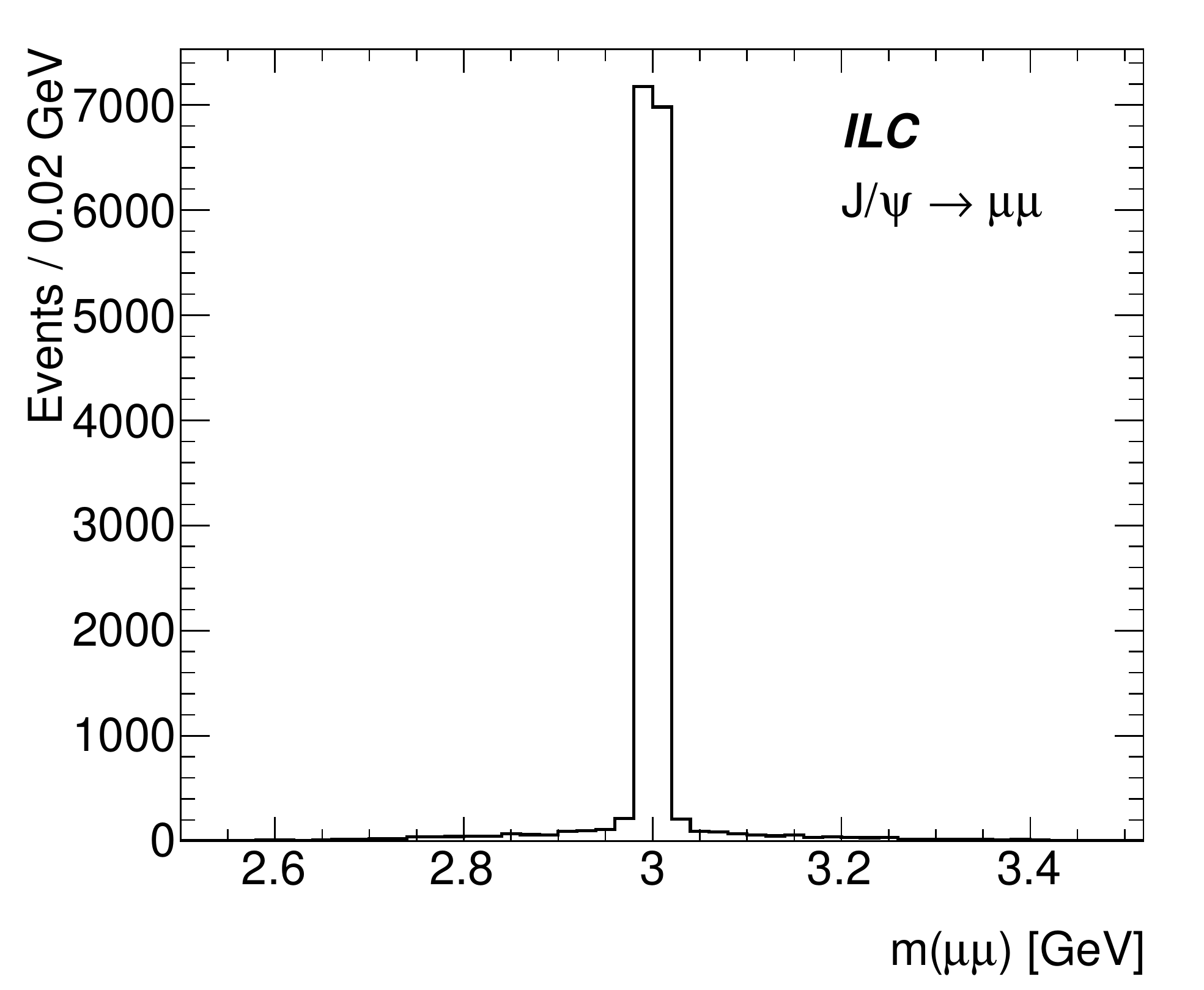}}\\
  \subfloat[]{
    \includegraphics[width=0.49\textwidth]{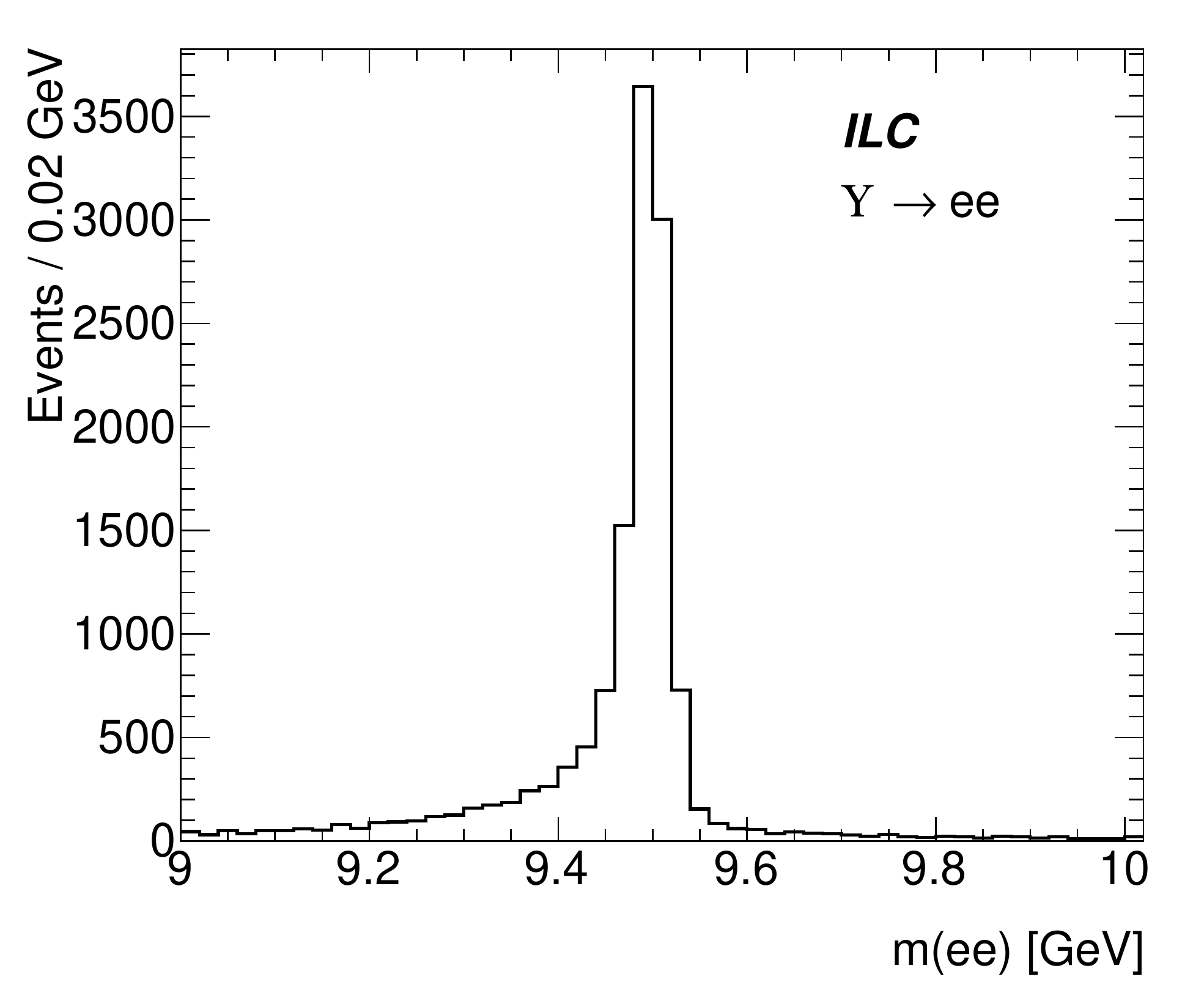}}
  \subfloat[]{
    \includegraphics[width=0.49\textwidth]{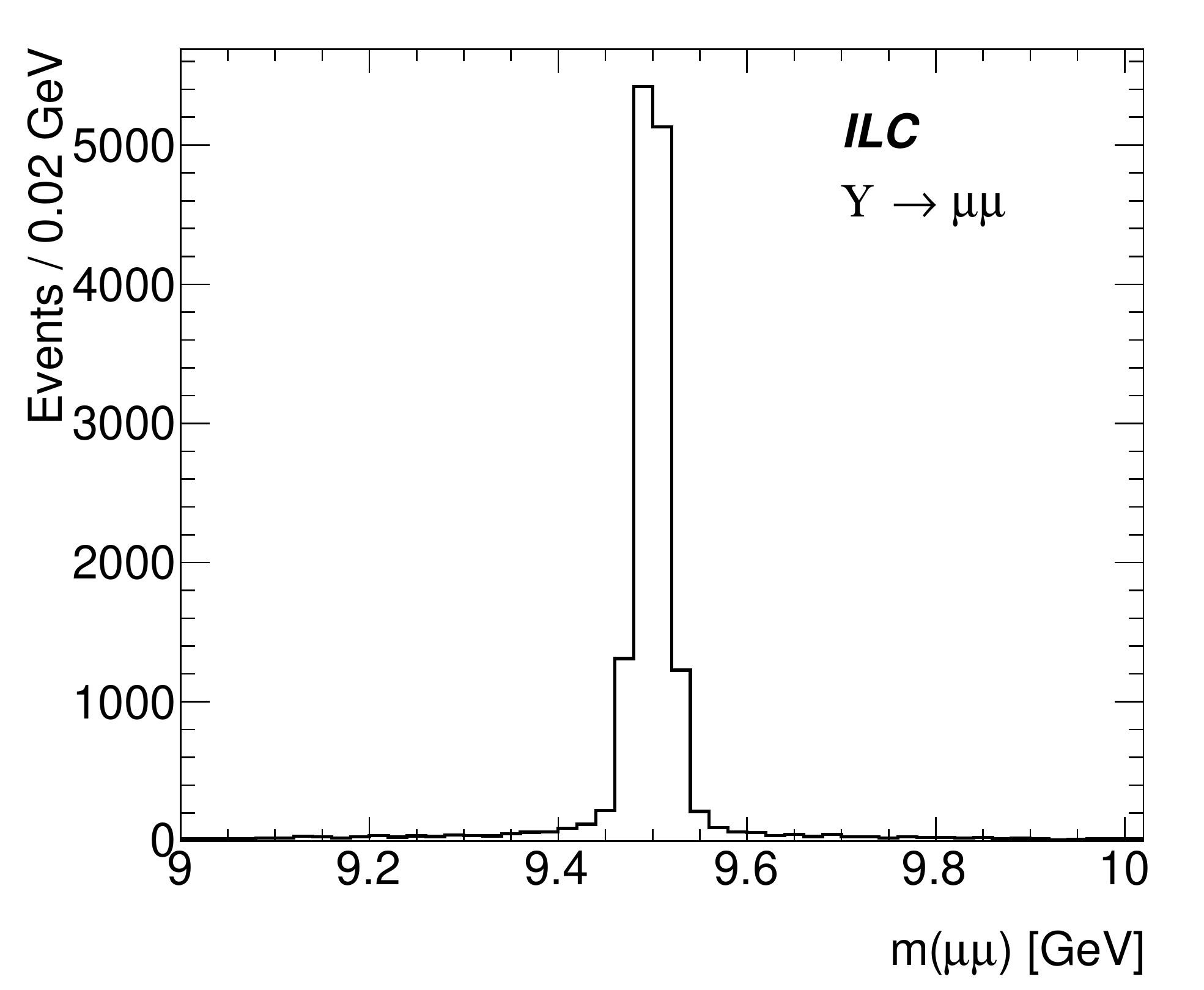}}\\
\end{center}
\caption{Invariant mass distributions for quarkonia decays to lepton pairs
  as generated by \HELAC+\PYTHIA:
  (a)~$J/\psi\rightarrow e^-e^+$;
  (b)~$J/\psi\rightarrow \mu^-\mu^+$;
  (c)~$\Upsilon\rightarrow e^-e^+$;
  (d)~$\Upsilon\rightarrow \mu^-\mu^+$.
  Note that \HELAC sets the quarkonia masses to be exactly the sum
  of the constituent quark masses; hence, the peaks are shifted from the
  true quarkonia masses.}
\label{fig:onia-peaks}
\end{figure}

Further, the invariant mass distributions of dilepton decays of quarkonia
are shown in \cref{fig:onia-peaks}.  These events were generated by
\HELAC+\PYTHIA and processed with the full detector simulation.
(\HELAC sets the quarkonia masses to be exactly the sum of the masses
of the constituent quarks, so the positions of the peaks are shifted
from the true quarkonia masses.)  These peaks are very narrow, especially
for the $\mu^-\mu^+$ decays.  Therefore, even if quarkonia backgrounds
were to be significant, they could be effectively suppressed by rejecting
a much smaller range in $m_{\Zd}$ than was done in the ATLAS analysis.

\section{\texorpdfstring{$H\rightarrow\Zd\Zd\rightarrow 4\ell$}{H->ZdZd->4l} expected limits}

Expected limits are set based on the distribution of the average
of the invariant masses of the two lepton pairs in a quadruplet,
$\left< m_{\ell\ell}\right> = \frac{1}{2} \left(m_{12} + m_{34}\right)$.
The likelihood function describing the data for a channel $j$
consists of a Poisson factor for each histogram bin $i$:
\begin{equation}
  {\cal L}(N) = \prod_i \textrm{Pois} \left(N_{ij}; \mu S_{ij}(m_{\Zd}) + B_{ij}\right), \label{eq:likelihoodDefinition1}
\end{equation}
where $S$ and $B$ are the predicted numbers of signal and background
events for each bin and channel and $\mu$ is the signal strength.
For this study, systematic uncertainties are assumed to be negligible.
The signal shape as a function of $m_{\Zd}$ is found by fitting a Gaussian
to the simulated signal at each generated mass point and then interpolating
in the fit mean and width as a function of $m_{\Zd}$.  Background
histograms are smoothed using the \verb|RooKeysPdf|
class of \textsc{RooFit}~\cite{RooFit,Cranmer:2000du},
except that if there are less
then ten simulated events surviving in a channel, the background
is taken to be flat with respect to $m_{\Zd}$.

\begin{figure}
\begin{center}
  \subfloat[]{
    \includegraphics[width=0.49\textwidth]{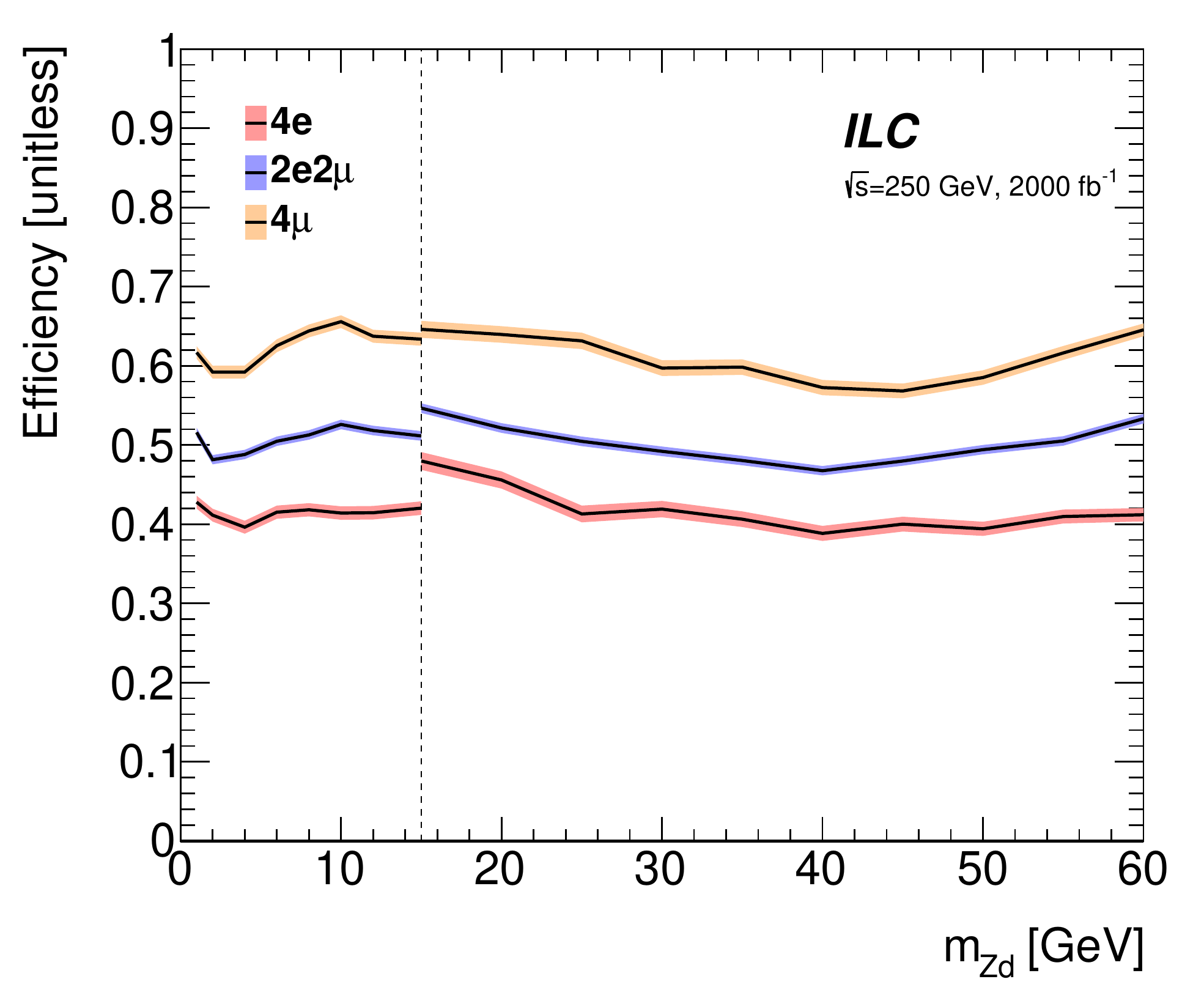}%
    \label{fig:4l-efficiency}}%
  \subfloat[]{
    \includegraphics[width=0.49\textwidth]{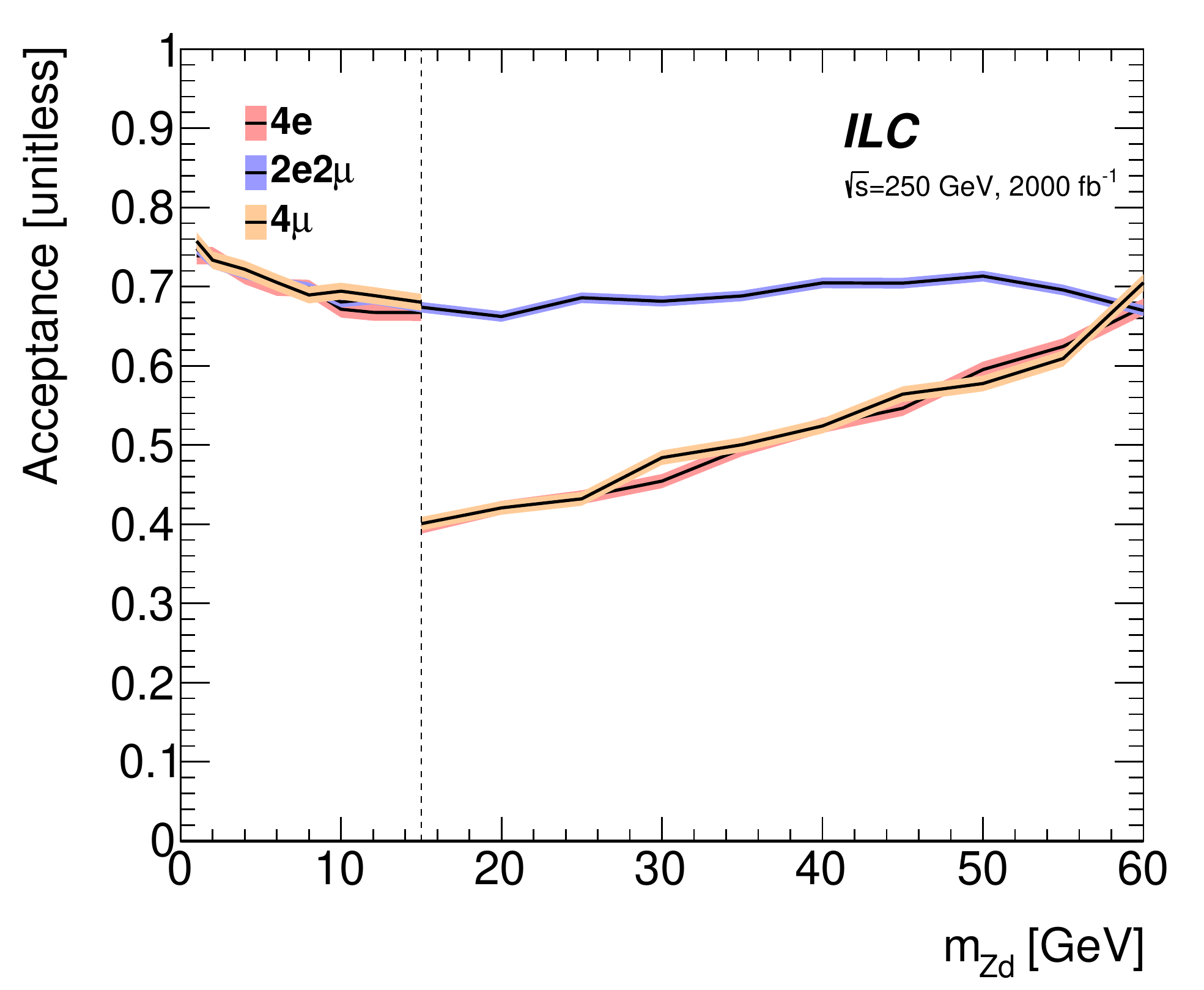}%
    \label{fig:4l-acceptance}}%
\end{center}
\caption{(a) Model-independent per-channel efficiencies for the fiducial
  volumes described in \cref{tab:4lfiducial}.  (b) Model-dependent per-channel
  acceptances for the $H\rightarrow \Zd\Zd\rightarrow 4\ell$ process.
  The discontinuities at $m_{\Zd} = \SI{15}{\GeV}$ are due to the change
  from the LM to HM selection.}
\label{fig:4l-effacc}
\end{figure}

Following~\cite{HDBS-2018-55}, a set of generator-level fiducial requirements,
described in \cref{tab:4lfiducial},
are used to factorize the event selection into a largely model-independent
`efficiency' and a model-dependent `acceptance'.  For the purpose of these
selections, the four-momenta of photons close to a lepton
($\Delta R < 0.1$) are added to that of the lepton.  This accounts
for the effects of quasi-collinear electromagnetic radiation
from the leptons~\cite{ATL-PHYS-PUB-2015-013}.  The efficiency for a channel
is defined
as the fraction of events passing the generator-level fiducial selection
that also passes the full event selection, while the acceptance
is defined as the fraction of generator-level events that pass the
fiducial selection.  The efficiency and acceptance for the analyses
described here are shown in \cref{fig:4l-effacc}.
For the HM selection, the acceptance falls for low $m_{\Zd}$
for the $4e$ and $4\mu$ channels due to the alternate pair requirement
of the $Z$-veto.
(Similar behavior was seen in the ATLAS analysis~\cite{HDBS-2018-55}.)

\begin{table}
  \caption{Summary of the fiducial phase-space definitions
    for the HM and LM analyses.  Objects are considered at generator-level,
    with photons nearby leptons summed with those leptons.
}
\centering
\small
\vskip 1em
\begin{tabular}{|c|>{\centering}m{5cm}|>{\centering}m{5cm}|}
  \cline{2-3}
  \multicolumn{1}{c|}{}
  & High-mass (HM) analysis \bigstrut[t]\\
    $H\rightarrow \Zd\Zd\rightarrow4\ell$ $(\ell=e,\mu)$ \bigstrut[b]
  & Low-mass (LM) analysis\\
    $H\rightarrow \Zd\Zd\rightarrow4\ell$ $(\ell=e,\mu)$ \tabularnewline
\hline

Mass range\bigstrut
  & $\SI{15}{\GeV} < m_X < \SI{60}{\GeV}$
  & $\SI{1}{\GeV} < m_X < \SI{15}{\GeV}$\tabularnewline
\hline

Leptons\bigstrut
  & \multicolumn{2}{>{\centering}m{10cm}|}{$E > \SI{7}{\GeV}$ and $0.35 < \theta < \pi - 0.35$}
  \tabularnewline
\hline

Quadruplet
  & \multicolumn{2}{>{\centering}m{10cm}|}{%
    Three leading-$E$ leptons satisfying $E > \SI{20}{\GeV}$, $\SI{15}{\GeV}$, $\SI{10}{\GeV}$\bigstrut} \tabularnewline
\cline{2-3}

\multicolumn{1}{|c|}{}
  & \multicolumn{1}{>{\centering}m{5cm}|}{\bigstrut[t] $\Delta R(\ell,\ell^{\prime})>0.10\,(0.20)$ for same-flavor (different-flavor) $\ell,\ell^{\prime}$\bigstrut}
  & ---\bigstrut \tabularnewline
\cline{2-3}

\multicolumn{1}{|c|}{}
  & \multicolumn{2}{>{\centering}m{10cm}|}{$m_{34}/m_{12} > 0.85$}\bigstrut
  \tabularnewline
\cline{2-3}

\multicolumn{1}{|c|}{}
  & $\SI{10}{\GeV} < m_{12,34} < \SI{64}{\GeV}$\bigstrut[t] \\
    For $4e$ and $4\mu$ channels: $\SI{5}{\GeV} < m_{14,23} < \SI{75}{\GeV}$ \bigstrut[b]
   & $\SI{0.8}{\GeV} < m_{12,34} < \SI{20}{\GeV}$
  \tabularnewline
\cline{2-3}

\multicolumn{1}{|c|}{}
  & \multicolumn{1}{>{\centering}m{5cm}|}{Reject event if $m_{12,34,14,23}$ in:\bigstrut[t]\\
  $(m_{J/\psi}-\SI{0.25}{\GeV})$ to $(m_{\psi(2S)}+\SI{0.30}{\GeV})$, or\\
  $(m_{\Upsilon(1S)}-\SI{0.70}{\GeV})$ to $(m_{\Upsilon(3S)}+\SI{0.75}{\GeV})$}\bigstrut[b]
  & \multicolumn{1}{>{\centering}m{5cm}|}{Reject event if $m_{14,23}$ in:\bigstrut[t]\\
  $(m_{J/\psi}-\SI{0.25}{\GeV})$ to $(m_{\psi(2S)}+\SI{0.30}{\GeV})$, or\\
  $(m_{\Upsilon(1S)}-\SI{0.70}{\GeV})$ to $(m_{\Upsilon(3S)}+\SI{0.75}{\GeV})$}\bigstrut[b]
\tabularnewline
  \hline
\end{tabular}
\label{tab:4lfiducial}
\end{table}

\begin{figure}
\begin{center}
  \subfloat[]{
    \includegraphics[width=0.49\textwidth]{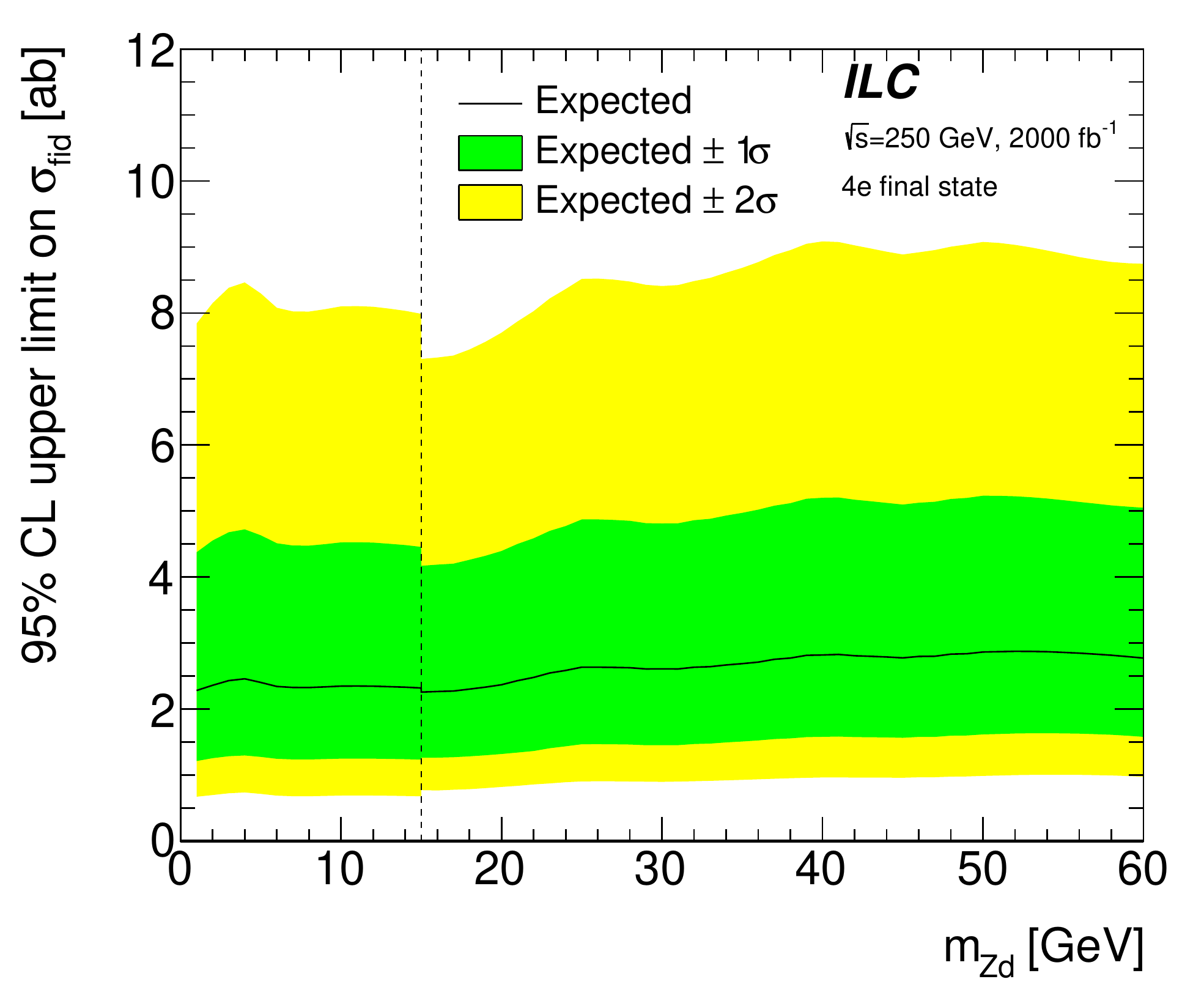}%
    \label{fig:4l-fid-eeee}}%
  \subfloat[]{
    \includegraphics[width=0.49\textwidth]{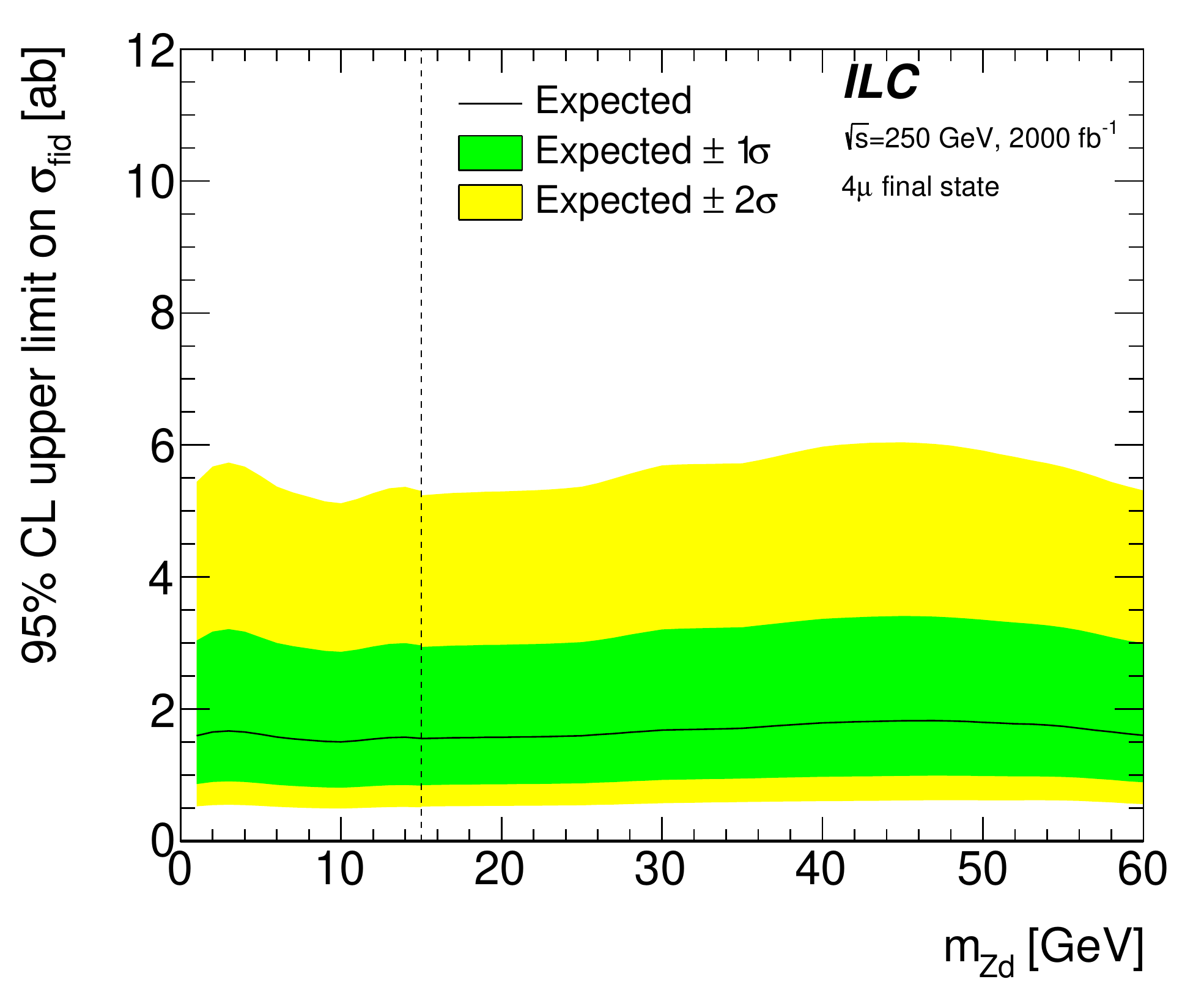}%
    \label{fig:4l-fid-mmmm}}\\%
  \subfloat[]{
    \includegraphics[width=0.49\textwidth]{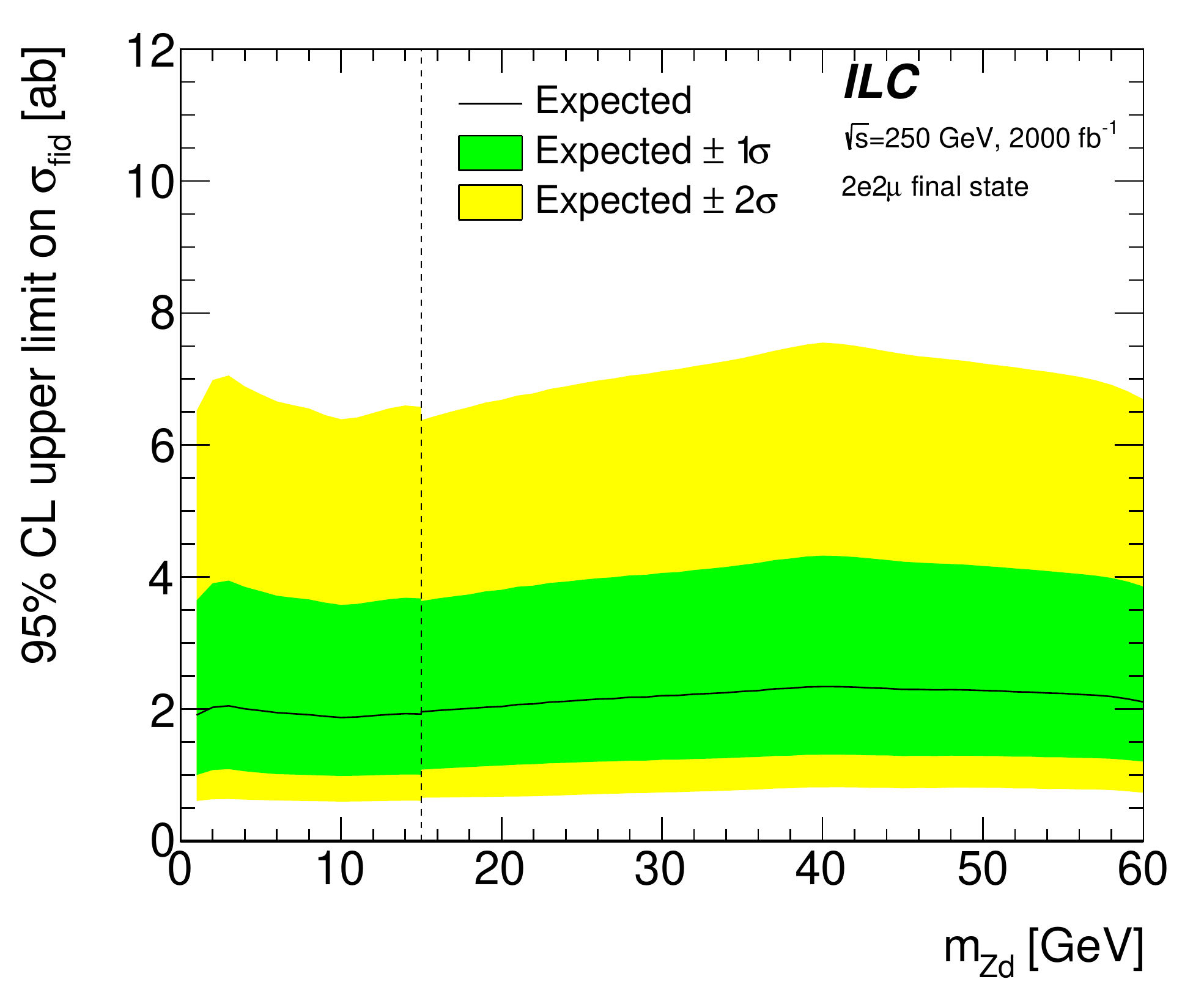}%
    \label{fig:4l-fid-eemm}}%
\end{center}
\caption{Per-channel expected upper limits at 95\% CL on the fiducial
  cross sections for the $H\rightarrow \Zd\Zd\rightarrow 4\ell$ process,
  for the (a) $4e$, (b) $4\mu$, and (c) $2e2\mu$ final states.
  The discontinuities at $m_{\Zd} = \SI{15}{\GeV}$ are due to the change
  from the LM to HM selection.}
\label{fig:4l-fid}
\end{figure}

\begin{figure}
\begin{center}
  \subfloat[]{
    \includegraphics[width=0.49\textwidth]{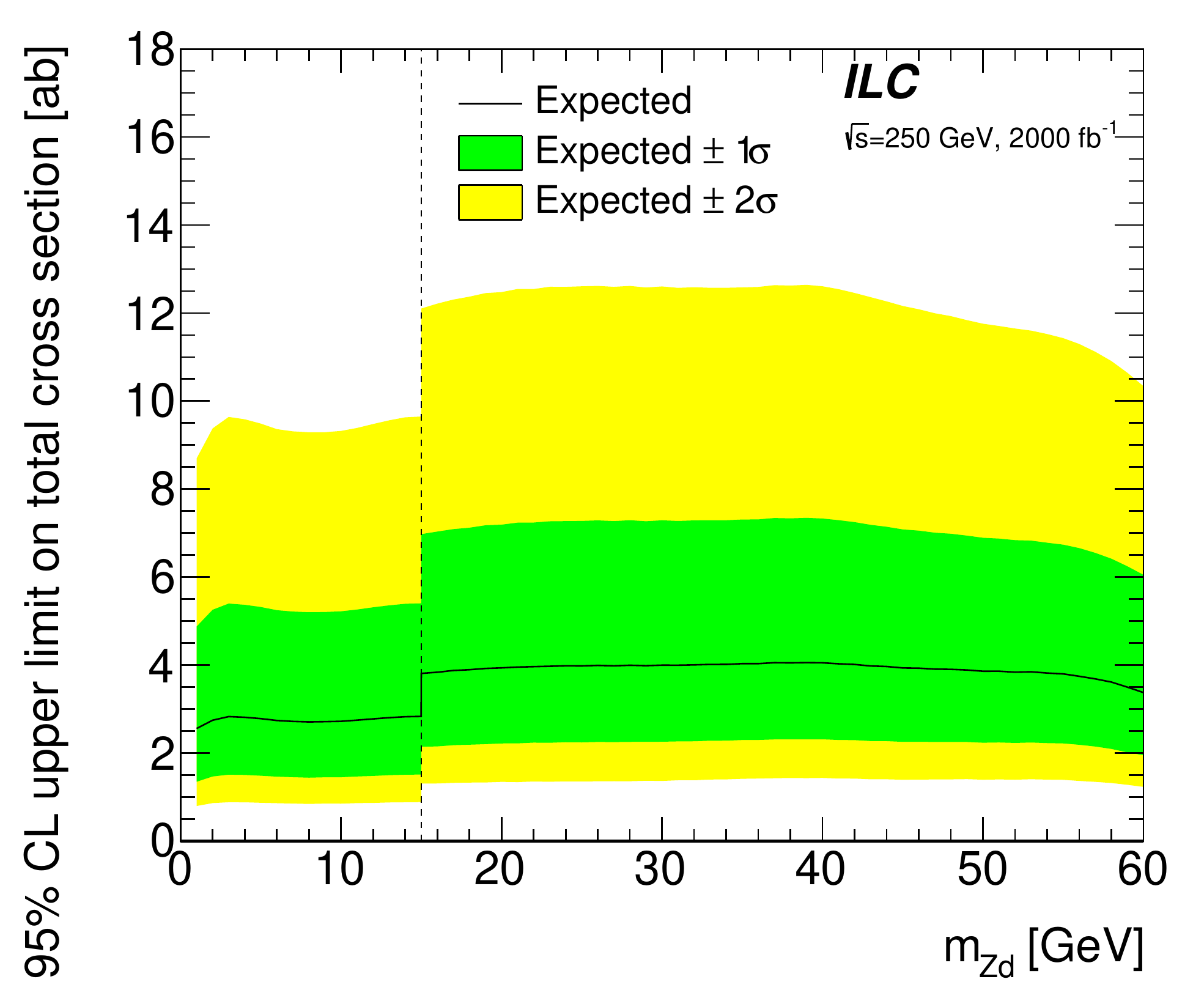}%
    \label{fig:4l-xstot}}%
  \subfloat[]{
    \includegraphics[width=0.49\textwidth]{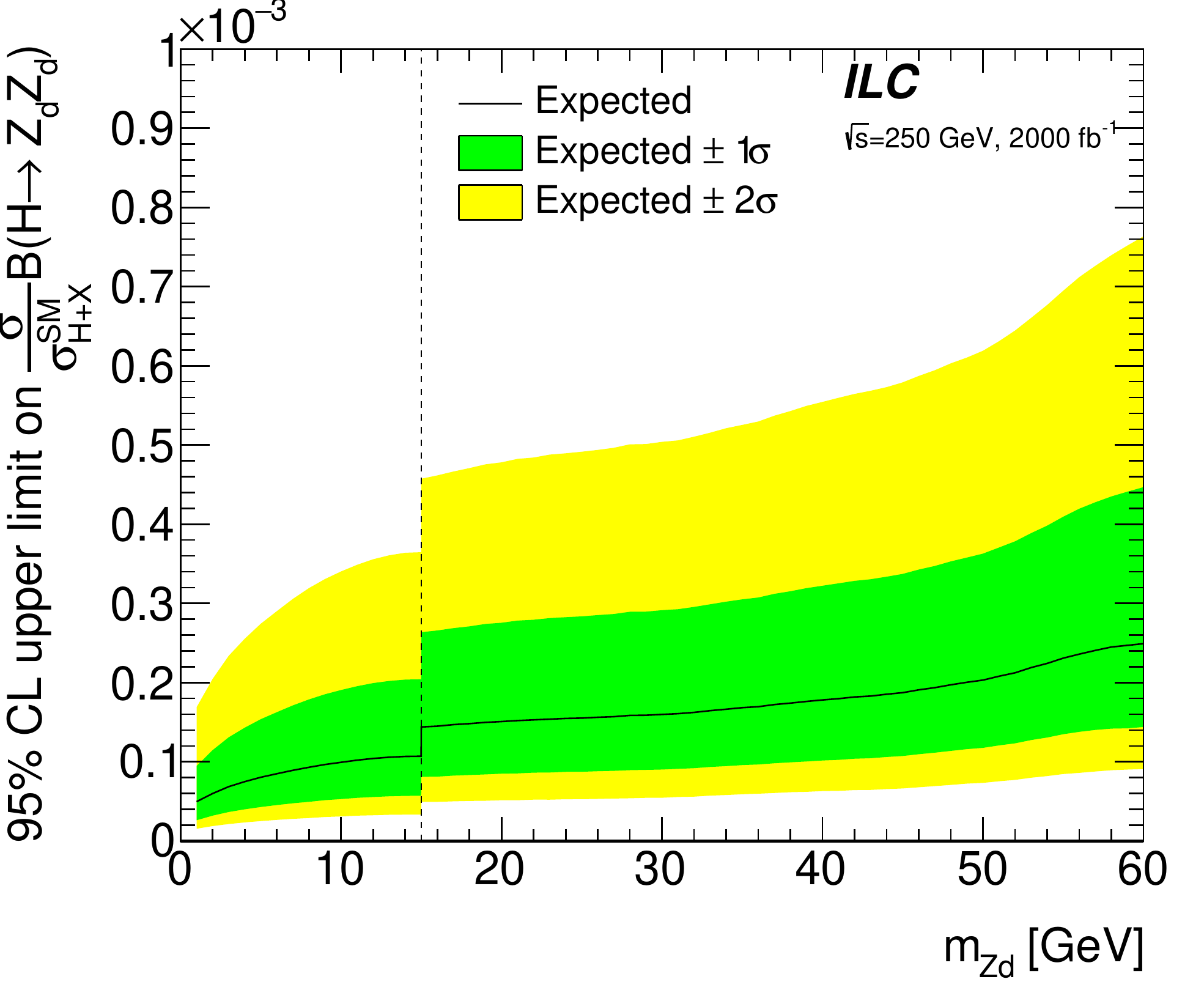}%
    \label{fig:4l-br}}%
\end{center}
\caption{(a) Expected upper limit at 95\% CL for the cross section
  of the $e^+e^- \rightarrow H+X \rightarrow \Zd\Zd+X \rightarrow 4\ell+X$
  process, assuming SM Higgs boson production.  (b) Expected upper limit
  at 95\% CL for the cross section times the model-dependent
  branching ratio divided by the SM Higgs boson production cross section
  (\SI{319}{fb}) for the $H\rightarrow \Zd\Zd$ process for the
  benchmark HAHM.  In both cases, all final states are combined.
  The discontinuities at $m_{\Zd} = \SI{15}{\GeV}$ are due to the change
  from the LM to HM selection.}
\label{fig:4l-limits}
\end{figure}

The efficiencies are used to compute expected 95\% CL upper limits
on the cross sections within the fiducial region,
using the CL$_\text{s}$ frequentist
formalism~\cite{Read:2002hq} with the profile-likelihood-ratio
test statistic~\cite{Cowan:2010js-witherratum},
and are shown in \cref{fig:4l-fid},
assuming a total integrated luminosity of $\SI{2000}{\ifb}$.
Incorporating the acceptance and combining the channels, this can be
converted into an upper limit on the product of the total cross
section and the decay branching ratio for the model considered,
$\sigma(e^+e^-\rightarrow H+X \rightarrow \Zd\Zd+X\rightarrow 4\ell + X)$, shown
in \cref{fig:4l-xstot}.  Using the model-dependent branching ratio
$\BR(\Zd\rightarrow 2\ell)$, this can be converted into an limit
on $\BR(H\rightarrow \Zd\Zd)$, shown in \cref{fig:4l-br}.

Compared to limits from the similar ATLAS analysis with
$\SI{139}{\ifb}$ of data and $\sqrt{s}=\SI{13}{\TeV}$,
the expected branching ratio limits here are a factor of 5--10 higher.
This is not unexpected: since the background is quite small,
even at the LHC, the sensitivity is driven mainly by the total number
of Higgs bosons produced, which was about ten times larger
at the LHC than would be expected in $\SI{2000}{\ifb}$ of ILC
data at $\sqrt{s}=\SI{250}{\GeV}$.  The exception is in the mass ranges
$\SI{2}{\GeV} < m_{\Zd} < \SI{4.4}{\GeV}$ and
$\SI{8}{\GeV} < m_{\Zd} < \SI{12}{\GeV}$, where the ATLAS analysis has no
sensitivity due to quarkonia backgrounds.

\section{\texorpdfstring{$H\rightarrow\Zd\Zd\rightarrow 2\ell2j$, $4j$}{H->ZdZd->2l2j, 4j} event selection}

The event selection for the $2\ell2j$ and $4j$ final states proceeds in two steps, 
beginning with a cut-based preselection followed by a selection based
on boosted decision trees (BDT).

The $2\ell2j$ final state requires at least one pair of opposite-sign 
electrons or muons along with four jets formed from
particle flow objects (PFO) not
associated with isolated leptons or photons. If there are multiple eligible 
lepton pairs, the one with invariant mass closest to the $Z$-boson mass 
$m_Z$ is selected as the $Z$ or $Z_d$ candidate. The $4j$ final state has no 
requirement on the number of isolated leptons, but requires six jets from 
non-isolated PFO objects. 
Either case uses the standard jet reconstruction to four or six jets,
respectively.

For both final states, all possible jet pairs are constructed. 
The invariant masses of the jet pairs, along with the dilepton's 
invariant mass in the $2\ell2j$ final state, are compared against $m_Z$ 
and the pair with the invariant mass closest to $m_Z$ is selected 
as the $Z$-boson candidate. 
If in the $2\ell2j$ final state preselection a jet pair is selected as  the
$Z$-candidate, the remaining jet and lepton pairs are selected as the
$Z_d$ candidates.  In all other cases all possible jet pairs from the 
remaining four jets are constructed and the two jet pairs 
with invariant mass closest to each other, 
i.e. minimizing $| m_{j_1j_2}-m_{j_3j_4} |$, are selected as the two $Z_d$ candidates.

The jets, as well as leptons in the $2\ell2j$ case, must satisfy the requirement on their polar angle $|\cos(\theta)|<0.9$.

The final requirement for the preselection is that the four fermions
constituting the two $Z_d$ candidates have a total invariant mass
broadly consistent with that of the Higgs~boson:
$\SI{90}{\GeV}<m_{ffff}<\SI{160}{\GeV}$.

The preselection efficiency $\times$ acceptance for the signal samples is around 31\% for the 
$4j$ final state and 24\% for the $2\ell2j$ one. For the background samples the 
efficiencies $\times$ acceptances are 17\% and 2\% for the $4j$ and $2\ell2j$ 
final states, respectively.

The signal regions are defined by boosted decision trees individually 
trained for each final state.  Half of the generated events
for both the background and signal samples
are randomly assigned for training, while the remainder are used 
for evaluation. The input variables for the BDT are:
\begin{itemize}
\item The transverse momentum, total energy, invariant mass, and $\cos(\theta)$ for each $Z$-boson and $Z_d$-boson candidate;
\item the $\Delta R$ between each possible boson candidate pair;
\item and the transverse momentum, total energy, invariant mass of the Higgs boson candidate.
\end{itemize}
The efficiencies after the preselection and the BDT selection for
the signals and background are shown in \cref{tab:4j_2l2j_selection_efficiency},
and \cref{tab:4j_2l2j_SR_yields} shows the signal region event yields assuming
an integrated luminosity of $\SI{2000}{\ifb}$ and an
$H\rightarrow\Zd\Zd\rightarrow 2\ell2j$, $4j$ cross section of \SI{1}{\femto\barn}.

\begin{table}
  \caption{Selection efficiencies $\times$ acceptance for the $4j$ and $2\ell2j$ signal regions after preselection and boosted decision tree selection.}
\centering
\small
\vskip 1em
\begin{tabular}{rcccc}
  \hline
  \bigstrut
         & \multicolumn{3}{c}{signal efficiency $\times$ acceptance } & background efficiency $\times$ acceptance\\
  \hline
  \bigstrut[t]
$m_{Z_d}$            &  \SI{20}{\GeV} & \SI{40}{\GeV}&\SI{60}{\GeV}&   \\
$4j$ final state     &  9.4\% & 5.2\% & 4.7 \% &0.02\% \\
$2\ell2j$ final state&   24\% & 22\%  & 24\% & 0.6\%  \\
\hline
\end{tabular}
\label{tab:4j_2l2j_selection_efficiency}
\end{table}

\begin{table}
  \caption{Expected signal region event yields for the $2\ell2j$ and $4j$
    final states. Both are scaled to an integrated luminosity of $\SI{2000}{\ifb}$.
    The signal yields assume a $H\rightarrow\Zd\Zd\rightarrow 2\ell2j$, $4j$ cross section of \SI{1}{\femto\barn}.}
\centering
\small
\vskip 1em
\begin{tabular}{rcccc}
  \hline
  \bigstrut
         & \multicolumn{3}{c}{signal yields } & background yield\\
  \hline
  \bigstrut[t]
$m_{Z_d}$            &  \SI{20}{\GeV} & \SI{40}{\GeV}&\SI{60}{\GeV}&   \\
$4j$ final state     &  187           & 103          &  93         &  8400 \\
$2\ell2j$ final state&  484           & 448          & 487         &  131 \\
\hline
\end{tabular}
\label{tab:4j_2l2j_SR_yields}
\end{table}

\section{\texorpdfstring{$H\rightarrow\Zd\Zd\rightarrow 2\ell2j$, $4j$}{H->ZdZd->2l2j, 4l}  expected limits}

\begin{figure}
\begin{center}
  \subfloat[]{
    \includegraphics[width=0.49\textwidth]{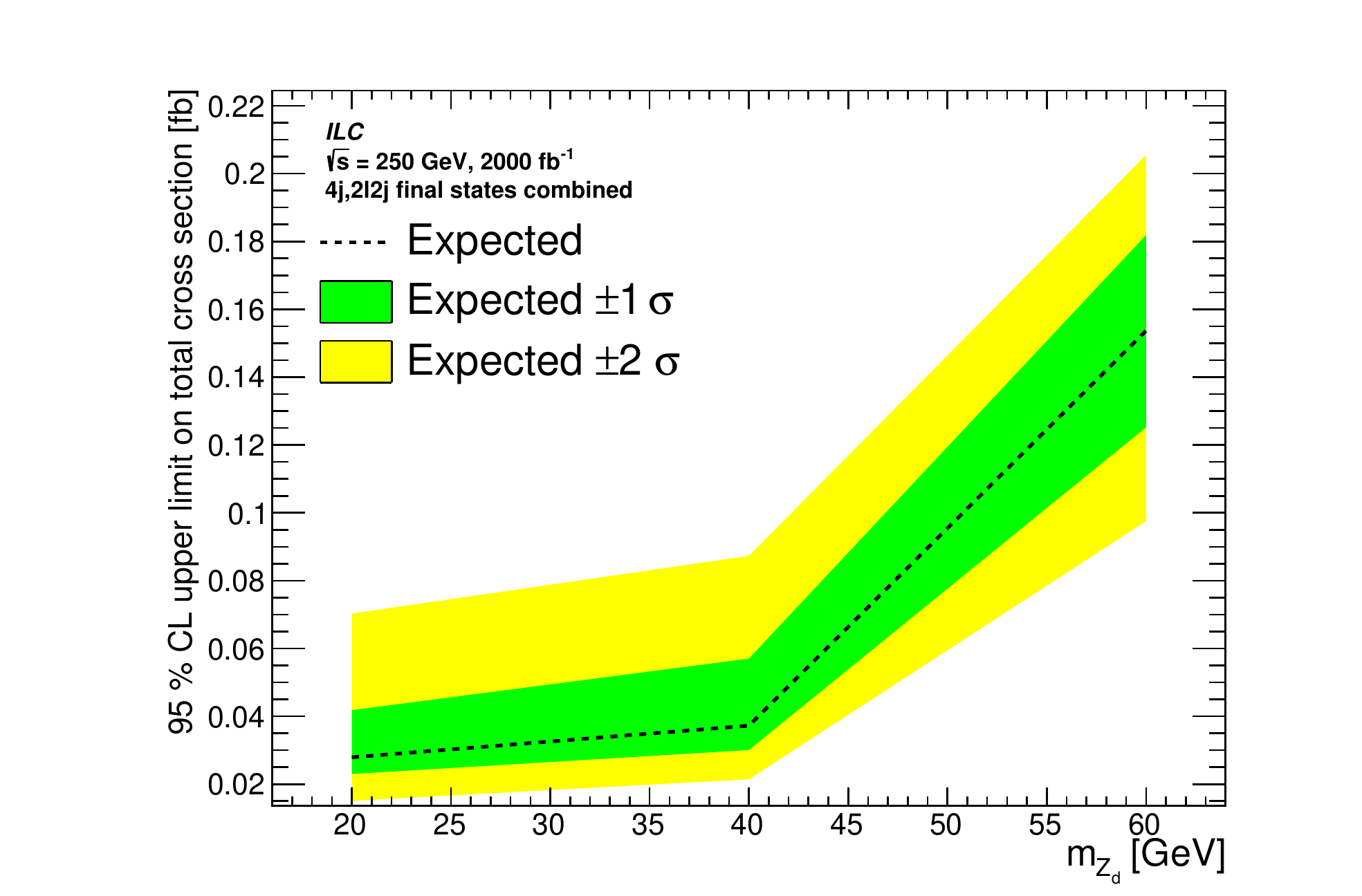}%
    \label{fig:jetFinalState-xstot}}%
  \subfloat[]{
    \includegraphics[width=0.49\textwidth]{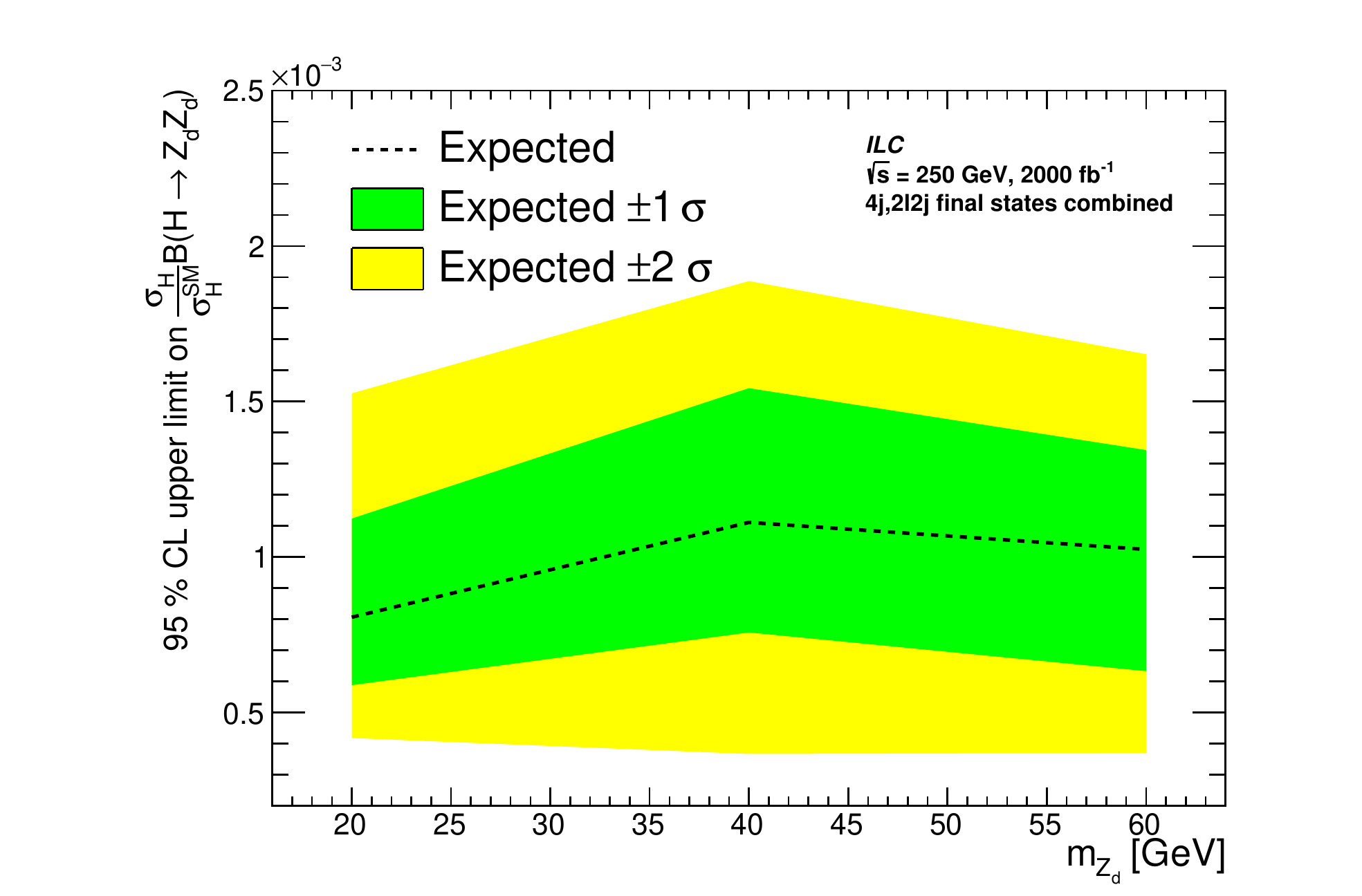}%
    \label{fig:jetFinalState-br}}%
\end{center}
\caption{(a) Expected upper limit at 95\% CL for the cross section
  of the $e^+e^- \rightarrow H+Z \rightarrow \Zd\Zd+Z \rightarrow 4j,2\ell2j+Z$
  process, assuming SM Higgs boson production. (b) Expected upper limit at 95\% CL on the $H \rightarrow \Zd\Zd$ branching ratio, derived from the $e^+e^- \rightarrow H+Z \rightarrow \Zd\Zd+Z \rightarrow 4j,2\ell2j+Z$ cross section. }
\label{fig:XSBRLimit_4j_2l2j}
\end{figure}

The likelihood function describing the data for the $4j$ and $2\ell2j$ final states follows \cref{eq:likelihoodDefinition1}. Both final state channels are fitted concurrently so the model in this case is given by:
\begin{equation}
  {\cal L}(N) = \prod_{j = 4j\text{,} 2l2j} \prod_i \textrm{Pois} \left(N_{ij}; \mu S_{ij}(m_{\Zd}) + B_{ij}\right). \label{eq:likelihoodDefinition2}
\end{equation}
The distributions used to evaluate the likelihood are the average
$\Zd$ mass: $\left< m_{Z_d} \right> = \frac{1}{2} \left(m_{Z_\text{d1}}+m_{Z_\text{d2}} \right)$.
The limits on the total cross section for the combined
$4j$ and $2\ell2j$ 
branching ratio are shown in \cref{fig:jetFinalState-xstot}. 

\cref{fig:jetFinalState-br} shows that the $2\ell2j$ and $4j$ final states do not yield a stricter expected limit on $\BR( H \to \Zd\Zd )$ than the $4\ell$ final state, despite the more favorable $\BR( \Zd \to 2j )$ branching ratio~\cite{Curtin:2013fra}.

\section{Summary and future work}

Expected limits have been presented for a search for dark photons
in the $H\rightarrow \Zd\Zd\rightarrow 4\ell$, $2\ell2j$, and $4j$ final states.  It is seen
that compared to the LHC, searches for these channels are not competitive
at the ILC, except for $\Zd$ masses close to the $J/\psi$ and $\Upsilon$
quarkonia resonances.  The LHC analysis does not have sensitivity in those
region due to large hadronic backgrounds, but it should be possible
to derive limits in those regions at the ILC where such backgrounds
are much smaller.  Doing this reliably, however, will likely require
progress in theoretical calculations of decays involving quarkonia states.

\section*{Acknowledgments}
This work is supported in part by the U.S.~Department of Energy under
contract DE-AC02-98CH10886 with Brookhaven National Laboratory.

\printbibliography


\end{document}
